\documentclass[prd,showpacs,twocolumn,superscriptaddress,floatfix,10pt]{revtex4-2}
\usepackage{setspace} 
\usepackage[utf8]{inputenc}
\usepackage{float}
\usepackage{amsmath,amssymb,amsfonts,bm}
\usepackage{graphicx}
\usepackage{multirow}
\usepackage[usenames,dvipsnames]{color}
\allowdisplaybreaks
\usepackage[
colorlinks=true,
linkcolor=blue,
breaklinks=true,
urlcolor=blue,
citecolor=blue]{hyperref}

\usepackage{orcidlink}
\usepackage{subfigure}
\usepackage{soul}
\usepackage{xcolor}

\definecolor{green}{rgb}{0,0.6,0}
\newcommand{\bra}[1]{\left\langle #1 \right|}
\newcommand{\ket}[1]{\left| #1 \right\rangle}

\newcommand{\be}{\begin{equation}} 
\newcommand{\ee}{\end{equation}}
\newcommand{\bea}{\begin{eqnarray}} 
\newcommand{\eea}{\end{eqnarray}}
\newcommand{\beas}{\begin{eqnarray*}} 
\newcommand{\eeas}{\end{eqnarray*}}

\newcommand{\mev}{\textrm{ MeV}}

\newcommand{\GXNU}{\affiliation{Department of Physics, Guangxi Normal University, Guilin 541004, China}}

\newcommand{\GXZD}{\affiliation{Guangxi Key Laboratory of Nuclear Physics and Technology, Guangxi Normal University, Guilin 541004, China}}

\newcommand{\IFIC}{\affiliation{Departamento de F\'{\i}sica Te\'orica and IFIC, Centro Mixto Universidad de
Valencia-CSIC Institutos de Investigaci\'on de Paterna, Apartado 22085,
46071 Valencia, Spain}}

\begin{document}
\title{The $\Omega_c \to \pi^+ \, (\pi^0,\, \eta)\, \pi \Xi^*$ reactions and the two $\Xi(1820)$ states}

\begin{abstract}
    We have studied the $\Omega_c \to \pi^+ (\pi^0, \eta) \pi \Xi^*$ decays, where the final $\pi \Xi^*$ comes from the decay of two resonances around the nominal $\Xi(1820)$, which are generated from the interaction of coupled channels made of a pseudoscalar and a baryon of the decuplet. The $\pi \Xi^*$ mass distributions obtained in the six different reactions studied are quite different and we single out four of them, which are free of a tree level contribution,  showing more clearly the effect of the resonances. The lower mass resonance is clearly seen as a sharp peak, but the higher mass resonance manifests itself through an interference with the lower one that leads to a dip in the mass distribution around $1850 \, {\rm MeV}$. Such a feature is similar to the dip observed in the $s$-wave $\pi \pi$ cross section around the $980 \, {\rm MeV}$ coming from the interference of the $f_0(500)$ and $f_0(980)$ resonances. Its observation in coming upgrades of present facilities will shed light on the existence of these two resonances and their nature.
\end{abstract}

\author{Wei-Hong Liang\orcidlink{0000-0001-5847-2498}}%
\email{liangwh@gxnu.edu.cn}
\GXNU%
\IFIC%
\GXZD%

\author{Raquel Molina\orcidlink{0000-0001-9427-240X}}
\email{raquel.molina@ific.uv.es}
\IFIC%

\author{Eulogio Oset\orcidlink{0000-0002-4462-7919}}%
\email{Oset@ific.uv.es}
\IFIC%
\GXNU%

\maketitle

\section{Introduction}

The issue of hadronic resonances corresponding to two nearby states with the same quantum numbers has been present for some time, since the challenging claim in Refs.~\cite{Oller:2000fj,Jido:2003cb} that the $\Lambda(1405)$ corresponded actually to two physical states.
In technical words, this means two nearby poles in the same Riemann sheet, not the shadow poles encountered in different Riemann sheets corresponding to the same physical state.
This issue was controversial at that time, but evidence from many theoretical calculations and different experiments opened the doors of the PDG \cite{PDG:2022} to the existence of
two $\Lambda(1405)$ states, and in the 2020 edition of the PDG \cite{PDG:2020} two $\Lambda(1405)$ states were officially admitted (see review paper on this issue \footnote{T. Hyodo and U.-G. Mei{\ss}ner, Pole Structure of the $\Lambda(1405)$ Region, Review paper in the PDG \cite{PDG:2022}.} on the PDG \cite{PDG:2022}).

The case of the two $\Lambda(1405)$ states opened the gates to the appearance of many other similar cases, one of them the two $K_1(1270)$ axial vector resonances, which were found in Ref.~\cite{Roca:2005nm}, and were supported experimentally as discussed in Ref.~\cite{Geng:2006yb}.
New cases were found for two $D^\ast_0(2400)$(now $D^*_0(2300)$) states in Refs.~\cite{Albaladejo:2016lbb,Gamermann:2006nm},
and, in the study of the $3/2^-$ baryons coming from the interaction of pseudoscalar mesons with baryons of the $3/2^+$ decuplet \cite{Kolomeitsev:2003kt,Sarkar:2004jh}, two states also emerged for the $\Xi(1820)$ resonance \cite{Sarkar:2004jh}.
There are also cases found from experimental analyzes, as the splitting of the $Y(4240)$ resonance reported by BaBar \cite{BaBar:2005hhc,BaBar:2006ait} into two states $Y(4230)$ and $Y(4260)$, suggested by the BESIII collaboration \cite{BESIII:2020bgb}.
In a recent paper \cite{Xie:2023cej} the authors show that the use of the Weinberg-Tomozawa interaction as the leading term of the chiral potentials  gives rise to a double pole structure of some states.
A global view on the issue of the double poles in hadronic resonances is presented in Ref.~\cite{Meissner:2020khl}.

Recently the BESIII collaboration reported on the reaction $\psi(3686) \to \bar \Xi^+ K^- \Lambda$ \cite{BESIII:2023mlv}, where an inspection of the 
$K^- \Lambda$ mass distribution showed two distinct peaks, one corresponding to the $\Xi(1690)$ resonance and another one associated to the $\Xi(1820)$, yet with a width ($\sim 73 \mev$) about three times bigger than the average width reported by the PDG \cite{PDG:2022} ($\sim 24 \mev$).
This apparent contradiction prompted a theoretical work \cite{Molina:2023uko}, where it was found that the apparent large width was a consequence of the contribution of the two $\Xi(1820)$ resonances.
Updating the work of Ref.~\cite{Sarkar:2004jh}, two poles were found in Ref.~\cite{Molina:2023uko}, one at $1824 \mev$ with a width of $62 \mev$, and a second one at $1875 \mev$ with a large width of $260 \mev$, and, with the contribution of the two states, a good reproduction of the BESIII mass distribution could be achieved.

In the present work, we look for an alternative reaction which can give information on the two $\Xi(1820)$ states.
The reaction is based on the weak decay of the $\Omega_c$ state and several decay channels are considered:
\begin{equation}\label{eq:reactions}
    \begin{split}
      \Omega_c &\to \pi^+ \,\Xi(1820) \to \pi^+ \pi^0 \,\Xi^{*-} \,(\pi^- \Xi^{*0}), \\[2mm]
      \Omega_c &\to \pi^0 \,\Xi(1820) \to \pi^0 \pi^+ \,\Xi^{*-}\,(\pi^0 \Xi^{*0}), \\[2mm]
      \Omega_c &\to \eta \,\Xi(1820) \to \eta \pi^+ \,\Xi^{*-}\,(\pi^0 \Xi^{*0}).
  \end{split} 
  \end{equation}
We find a clear contribution of the two states in all these reactions, but showing in a peculiar way, through a destructive interference that leads to a pronounced dip in the $\pi \Xi^*$ mass distribution around $1850 \mev$. 
This situation reminds one of the same features seen in the $\pi\pi$ isospin $I=0$, $s$-wave scattering, where the cross section has a broad peak corresponding to the $f_0(500)$ resonance and a dip corresponding to the $f_0(980)$ \cite{Pelaez:2015qba,Oller:1997yg} (see also this dip in the $0^{++} \pi^0 \pi^0$ mode in $J/\psi$ radiative decay to two pions \cite{BESIII:2015rug}).

We shall also see different shapes of the mass distributions for these reactions reflecting the presence of more than one resonance.
This was one of the experimental arguments used in favor of the two $\Lambda(1405)$ states by comparing the different shapes of the $\pi \Sigma$ mass distribution in the $\pi^- p \to K^0 \pi \Sigma$ \cite{Thomas:1973uh} and $K^- p \to \pi^0\pi^0 \Sigma^0$ \cite{CrystallBall:2004ovf} (see the discussion on this issue in Ref.~\cite{Magas:2005vu}).

\section{Formalism}

\subsection{The two $\Xi^*(1820)$ states}
In Ref.~\cite{Sarkar:2004jh}, the coupled channels of pseudoscalar meson-baryon ($3/2^+$) leading to baryons with strangeness $S=-2$ were considered, and with the interaction borrowed from chiral Lagrangians and a unitary scheme, two $\Xi^*$ states with $3/2^-$ emerged.
An update of the approach is done in Ref.~\cite{Molina:2023uko} and the amplitudes obtained there are used here.
The coupled channels are
\begin{equation*}
	\bar K^0\Sigma^{*-}, ~ K^-\Sigma^{*0},~ \pi^0 \Xi^{*-}, ~\eta \Xi^{*-},~ \pi^- \Xi^{*0},~  K^0\Omega^-,
\end{equation*}
with charge $Q=-1$, and 
\begin{equation*}
	\bar K^0\Sigma^{*0}, ~ K^-\Sigma^{*+},~ \pi^+ \Xi^{*-}, ~ \pi^0 \Xi^{*0},~ \eta \Xi^{*0},~ K^+\Omega^-,
\end{equation*}
with charge $Q=0$.
The interaction (potential) is given by 
\begin{equation}\label{eq:Vij}
	V_{ij}=-\dfrac{1}{4f^2} C_{ij} (k^0+k^{\prime \,0});~ f=1.28\, f_\pi,~ f_\pi =93 \mev,
\end{equation}
with $k^0, k^{\prime \,0}$ the energies of the pseudoscalar mesons, and $C_{ij}$ the coefficients given in tables A.4.2 and A.4.3 of Ref.~\cite{Sarkar:2004jh}.
The scattering matrix is obtained via the Bethe-Salpeter (BS) equation in matrix form 
\begin{equation}\label{eq:BS}
  T=[1-VG]^{-1} \, V,
\end{equation}
with $G$ the meson-baryon loop function, regularized with a cutoff $q_{\rm max}=830 \mev$, in order to get a good reproduction of the BESIII data \cite{BESIII:2023mlv,Molina:2023uko}. 

\subsection{Weak decay}
\begin{figure*}[!hbt]
    \centering
    \subfigure[]{\includegraphics[scale=0.6]{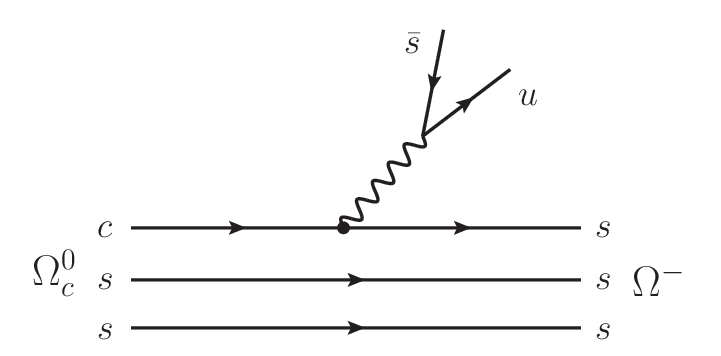}\label{fig:Fig1a}}
    \subfigure[]{\includegraphics[scale=0.6]{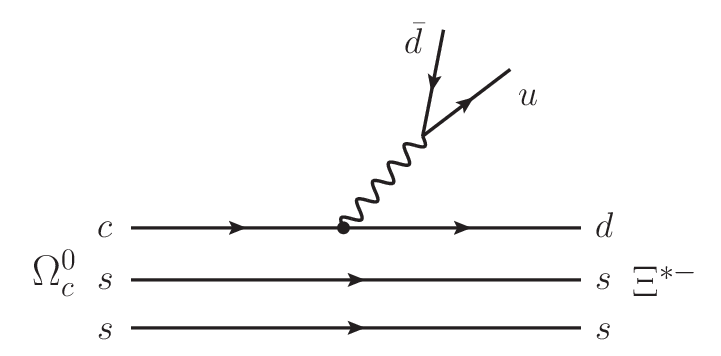}\label{fig:Fig1b}}
    \caption{The two topological structures with external emission that lead to $\Omega^-$ or $\Xi^{*-}$ in the final state.}
    \label{fig:Fig1}
\end{figure*}

The process that we study is single Cabibbo suppressed.
We consider the dominant external emission mechanism. 
At the quark level, we have two topologies that can lead to the desired final state depicted in Figs.~\ref{fig:Fig1a} and \ref{fig:Fig1b}.
In order to obtain two mesons in the final state, we have to hadronize the $\bar s u$ and $\bar d u$ components.
This is done by writing the matrix $q\bar q$ in terms of physical mesons, $P$,
\begin{eqnarray}\label{eq:Pmatrix}
    P &=& \left(\begin{array}{ccc}
        \frac{\pi^0}{\sqrt{2}}+\frac{\eta}{\sqrt{3}} & \pi^{+} & K^{+} \\
    \pi^{-} & -\frac{\pi^0}{\sqrt{2}}+\frac{\eta}{\sqrt{3}} & K^0  \\
    K^{-} & \bar{K}^0 & -\frac{\eta}{\sqrt{3}} \\
    \end{array}\right), 
\end{eqnarray}
where the $\eta$-$\eta'$ standard mixing of Ref.~\cite{Bramon:1992kr} is used, and the $\eta'$, not playing a role in the energy region of relevance, is omitted.
Then, 
\begin{eqnarray}\label{eq:usbar}
    u\bar s &\to&  \sum_i u\bar q_i q_i \bar s =P_{1i}\, P_{i3} =(P^2)_{13} \nonumber \\
    &=& \left(\frac{\pi^0}{\sqrt{2}}+\frac{\eta}{\sqrt{3}}\right)\, K^+ \, +\, \pi^+ K^0 \,-\, \frac{1}{\sqrt{3}}K^+ \eta.
\end{eqnarray}
It might look like the $\eta K^+$ component cancels, but this is not the case, as we see below, because the order matters.
\begin{eqnarray}\label{eq:udbar}
    u\bar d &\to&  \sum_i u\bar q_i q_i \bar d =P_{1i}\, P_{i2} =(P^2)_{12} \nonumber \\
    &=&\!\!\!(\frac{\pi^0}{\sqrt{2}}+\frac{\eta}{\sqrt{3}}) \pi^+  + \pi^+ (-\frac{\pi^0}{\sqrt{2}}+\frac{\eta}{\sqrt{3}}) + \! K^+ \bar K^0.
\end{eqnarray}
Once again, the $\pi^0\pi^+$ component does not cancel, but the $\eta \pi^+$ does, as we see below.

The coupling of $W^+$ to the meson-meson component has the structure of $\langle [P, \partial_\mu P]\, W^\mu \,T_-\rangle$, with $T_-$ a matrix related to the Kobayashi-Maskawa elements \cite{Gasser:1983yg,Scherer:2002tk}.
The $csW$ vertex is of the type $\gamma^\mu (1-\gamma_5)$, and the resulting weak transition operator at the quark level is $(p_1-p_2)_\mu\, \gamma^\mu (1-\gamma_5)$ with $p_1, p_2$ the momenta of the first, second meson.
But we have to make a transition from a spin $1/2^+$ state ($\Omega^0_c$) to a $3/2^+$ state $\Omega^-$ or $\Xi^{*-}(1530)$, which requires a spin operator at the quark level and we need then the term $(p_1-p_2)^i \gamma^i \gamma_5 \to \sigma^i (p_1-p_2)^i$.
This operator at the macroscopic level between the $\Omega_c^0$ and the $\Omega^-, \Xi^{*-}$ states has the type
\begin{equation}\label{eq:operator}
    \langle \Omega^-\, (\Xi^{*-}(1530)) \big|\vec S^+ \cdot (\vec p_1- \vec p_2) \big|  \Omega_c^0 \rangle,
 \end{equation}
 where $S^+$ is the spin transition operator from spin $1/2$ to spin $3/2$, which has the property in cartesian basis
 \begin{equation}\label{eq:SM}
     \sum_M S_i \ket{M} \bra{M} S^+_j = \frac{2}{3} \delta_{ij} -\frac{i}{3} \epsilon_{ijk} \;\sigma_k.
 \end{equation}
 From this perspective, we see that the $\eta K^+$ and $-K^+ \eta$ term in Eq.~\eqref{eq:usbar} give the same contribution, and so do the $\pi^0\pi^+$ and $-\pi^+\pi^0$ of Eq.~\eqref{eq:udbar}, while the terms $\eta \pi^0, \pi^0 \eta$ of Eq.~\eqref{eq:udbar} cancel.

 The mechanisms of Figs.~\ref{fig:Fig1a} and \ref{fig:Fig1b} share the same Cabibbo strength factor $\cos\theta_c \sin\theta_c$, but the matrix elements are different. 
 Indeed, we have for the mechanism of Fig.~\ref{fig:Fig1a},
 \begin{eqnarray}\label{eq:fig1a}
      && \bra{sss \, \chi_S} \, \vec\sigma \cdot (\vec p_1-\vec p_2) \bar c s\ket{css \, \chi_{MS}}\nonumber \\
     &=& \bra{\chi_S} \vec\sigma \cdot (\vec p_1-\vec p_2) \ket{\chi_{MS}}.
 \end{eqnarray}
 %
 while for Fig.~\ref{fig:Fig1b} we have
 \begin{eqnarray}\label{eq:fig1b}
     && \langle\frac{1}{\sqrt{3}}(dss+sds+ssd) \, \chi_S |\, \vec\sigma \cdot (\vec p_1-\vec p_2) \bar c d |css \, \chi_{MS} \rangle\nonumber \\
    &=& \frac{1}{\sqrt{3}} \, \bra{\chi_S} \vec\sigma \cdot (\vec p_1-\vec p_2) \ket{\chi_{MS}},
 \end{eqnarray}
 where $\chi_S$ and $\chi_{MS}$ are the spin symmetric and mixed symmetric wave functions.
Hence we see that the two matrix elements have the same spin structure, but the flavor structure gives an extra factor $\frac{1}{\sqrt{3}}$ for the mechanism of Fig.~\ref{fig:Fig1b}.
Note that we take the $css \,\chi_{MS}$ structure for the $\Omega_c^0$ state, singling out the $c$ quark, following Refs.~\cite{Capstick:1986ter,Roberts:2007ni}.

\begin{figure}[t]
    \centering
    \includegraphics[scale=0.5]{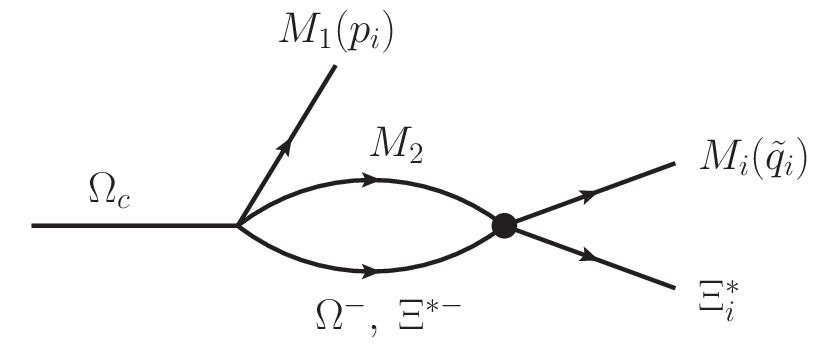}
    \caption{Final state interaction of a meson with the baryon of the decuplet $3/2^+$. The dot indicates the transition matrix element from $M_2 \Omega^-(\Xi^{*-})$ to a final $M_i \Xi^*_i$ state.}
    \label{fig:Fig2}
\end{figure}

In order to generate the $\Xi(1820)$ resonance in the final state, we have to allow one of the mesons to interact with the $\Omega^-$ or $\Xi^{*-}$ and this leads to the picture of Fig.~\ref{fig:Fig2}.
The structure of Fig.~\ref{fig:Fig2} corresponds to an amplitude of the type
\begin{eqnarray}\label{eq:ampli}
        && \int_{|\vec p_2\,| < q_{\rm max}} \dfrac{{\rm d}^3 p_2}{(2\pi)^3} \; \langle \Omega^- | \vec S^+ \cdot (\vec p_1 -\vec p_2) | \Omega_c^0 \rangle \;
        \nonumber\\[2mm]
         &\cdot&\dfrac{1}{2\,\omega(p_2)}\;\dfrac{m_B}{E_B(p_2)}\;\dfrac{1}{M_{\rm inv}-\omega(p_2)-E_B(p_2) + i \varepsilon} \nonumber\\[2mm]
         &\cdot& t_{M_2 B,\, M_i \Xi^*_i} (M_{\rm inv}(M_i \Xi^*_i)),
    \end{eqnarray}
    where $\omega(p_2)=\sqrt{\vec p_2^{\;2}+m_{M_2}^2}$, $E_B(p_2)=\sqrt{\vec p_2^{\;2}+m_{B}^2}$, with $M_2, B$ the intermediate meson, baryon states in the loop, and $t_{M_2 B,\, M_i \Xi^*_i}$ the transition scattering matrix from $M_2 B$ to $M_i \Xi^*_i$.
    Since $t_{M_2 B,\, M_i \Xi^*_i}$ is constructed with the $s$-wave potential of Eq.~\eqref{eq:Vij}, the term with $\vec p_2$ in Eq.~\eqref{eq:ampli} does not contribute, hence, only the $\vec S^+ \cdot \vec p_1$, corresponding to the external meson in the weak vertex, contributes.
    Then the spin matrix element of Eq.~\eqref{eq:ampli} factorizes out of the integral and so does $t_{M_2 B,\, M_i \Xi^*_i}$.
    All this said, we have six possible reactions written below, where the first meson corresponds to the external one of the weak vertex and the second one to the final state. 
    The corresponding amplitudes are written for each case.
    First we look at the terms originating from Fig.~\ref{fig:Fig1a}, with an $\Omega^-$ in the intermediate state,

    \vspace{0.2cm}
    1) $\Omega_c^0 \to \pi^+ \pi^0 \Xi^{*-}$ 
    \begin{eqnarray}\label{eq:t1}
      t_1 &=& C\; \langle \Omega^- | \vec S^+ \cdot \vec p_{\pi^+} | \Omega_c^0 \rangle\; t'_1, \\
      t'_1 &=& G_{K^0 \Omega^-} (M_{\rm inv}(\pi^0 \Xi^{*-})) \cdot t_{K^0 \Omega^-, \pi^0 \Xi^{*-}}(M_{\rm inv}(\pi^0 \Xi^{*-}));\nonumber
    \end{eqnarray}
    
    2) $\Omega_c^0 \to \pi^+ \pi^- \Xi^{*0}$ 
    \begin{eqnarray}\label{eq:t2}
      t_2 &=& C\; \langle \Omega^- | \vec S^+ \cdot \vec p_{\pi^+} | \Omega_c^0 \rangle\; t'_2, \\
      t'_2 &=& G_{K^0 \Omega^-} (M_{\rm inv}(\pi^- \Xi^{*0})) \cdot t_{K^0 \Omega^-, \pi^- \Xi^{*0}}(M_{\rm inv}(\pi^- \Xi^{*0}));\nonumber
    \end{eqnarray}

3) $\Omega_c^0 \to \pi^0 \pi^+ \Xi^{*-}$ 
\begin{eqnarray}\label{eq:t3}
  t_3 &=& C\; \langle \Omega^- | \vec S^+ \cdot \vec p_{\pi^0} | \Omega_c^0 \rangle\; t'_3, \\
  t'_3 &=& \frac{1}{\sqrt{2}}G_{K^+ \Omega^-} (M_{\rm inv}(\pi^+ \Xi^{*-})) \cdot t_{K^+ \Omega^-, \pi^+ \Xi^{*-}}(M_{\rm inv}(\pi^+ \Xi^{*-}));\nonumber
\end{eqnarray}

4) $\Omega_c^0 \to \pi^0 \pi^0 \Xi^{*0}$ 
\begin{eqnarray}\label{eq:t4}
  t_4 &=& C\; \langle \Omega^- | \vec S^+ \cdot \vec p_{\pi^0} | \Omega_c^0 \rangle\; t'_4, \\
  t'_4 &=& \frac{1}{\sqrt{2}}G_{K^+ \Omega^-} (M_{\rm inv}(\pi^0 \Xi^{*0})) \cdot t_{K^+ \Omega^-, \pi^0 \Xi^{*0}}(M_{\rm inv}(\pi^0 \Xi^{*0}));\nonumber
\end{eqnarray}

5) $\Omega_c^0 \to \eta \pi^+ \Xi^{*-}$ 
\begin{eqnarray}\label{eq:t5}
  t_5 &=& C\; \langle \Omega^- | \vec S^+ \cdot \vec p_{\eta} | \Omega_c^0 \rangle\; t'_5, \\
  t'_5 &=& \frac{2}{\sqrt{3}}G_{K^+ \Omega^-} (M_{\rm inv}(\pi^+ \Xi^{*-})) \cdot t_{K^+ \Omega^-, \pi^+ \Xi^{*-}}(M_{\rm inv}(\pi^+ \Xi^{*-}));\nonumber
\end{eqnarray}

6) $\Omega_c^0 \to \eta \pi^0 \Xi^{*0}$ 
\begin{eqnarray}\label{eq:t6}
  t_6 &=& C\; \langle \Omega^- | \vec S^+ \cdot \vec p_{\eta} | \Omega_c^0 \rangle\; t'_6, \\
  t'_6 &=& \frac{2}{\sqrt{3}}G_{K^+ \Omega^-} (M_{\rm inv}(\pi^0 \Xi^{*0})) \cdot t_{K^+ \Omega^-, \pi^0 \Xi^{*0}}(M_{\rm inv}(\pi^0 \Xi^{*0}));\nonumber
\end{eqnarray}
where $C$ is a normalization constant, common to all terms.

Next we look at the amplitudes stemming from Fig.~\ref{fig:Fig1b}, leading to a $\Xi^{*-}$ in the intermediate state.
We have

\vspace{0.2cm}
7) $\Omega_c^0 \to \pi^+ \pi^0 \Xi^{*-}$ 
\begin{eqnarray}\label{eq:t7}
  t_7 &=& C\; \langle \Xi^{*-} | \vec S^+ \cdot \vec p_{\pi^+} | \Omega_c^0 \rangle\; t'_7,\\
  t'_7 &=&-\sqrt{\frac{2}{3}}\left[ 1+ G_{\pi^0 \Xi^{*-}} (M_{\rm inv}(\pi^0 \Xi^{*-}))
  \right.\nonumber\\
  &&~~~~~~~~~~~~~\left. \cdot t_{\pi^0 \Xi^{*-}, \pi^0 \Xi^{*-}}(M_{\rm inv}(\pi^0 \Xi^{*-}))\right];\nonumber
\end{eqnarray}

8) $\Omega_c^0 \to \pi^+ \pi^- \Xi^{*0}$ 
\begin{eqnarray}\label{eq:t8}
  t_8 &=& C\; \langle \Xi^{*-} | \vec S^+ \cdot \vec p_{\pi^+} | \Omega_c^0 \rangle\; t'_8,\\
  t'_8 &=& -\sqrt{\frac{2}{3}}\, G_{\pi^0 \Xi^{*-}} (M_{\rm inv}(\pi^- \Xi^{*0})) \nonumber\\
  &&~~~~~~~\cdot t_{\pi^0 \Xi^{*-}, \pi^- \Xi^{*0}}(M_{\rm inv}(\pi^- \Xi^{*0}));\nonumber
\end{eqnarray}

9) $\Omega_c^0 \to \pi^0 \pi^+ \Xi^{*-}$ 
\begin{eqnarray}\label{eq:t9}
  t_9 &=& C\; \langle \Xi^{*-} | \vec S^+ \cdot \vec p_{\pi^0} | \Omega_c^0 \rangle\; t'_9,\\
  t'_9 &=&\sqrt{\frac{2}{3}}\left[ 1+ G_{\pi^+ \Xi^{*-}} (M_{\rm inv}(\pi^+ \Xi^{*-})) \right.\nonumber\\
  &&~~~~~~~~~~~~\left.\cdot t_{\pi^+ \Xi^{*-}, \pi^+ \Xi^{*-}}(M_{\rm inv}(\pi^+ \Xi^{*-}))\right];\nonumber
\end{eqnarray}

10) $\Omega_c^0 \to \pi^0 \pi^0 \Xi^{*0}$ 
\begin{eqnarray}\label{eq:t10}
  t_{10} &=& C\; \langle \Xi^{*-} | \vec S^+ \cdot \vec p_{\pi^0} | \Omega_c^0 \rangle\; t'_{10},\\
  t'_{10} &=& \sqrt{\frac{2}{3}}\, G_{\pi^+ \Xi^{*-}} (M_{\rm inv}(\pi^0 \Xi^{*0})) 
  \nonumber\\
  &&~~~~~\cdot t_{\pi^+ \Xi^{*-}, \pi^0 \Xi^{*0}}(M_{\rm inv}(\pi^0 \Xi^{*0}));\nonumber
\end{eqnarray}

The cases 7) - 10) in Eqs.~\eqref{eq:t7}-\eqref{eq:t10} correspond to the same final state than
for the cases 1) -4) and have to be added coherently.
In the amplitudes of Eqs.~\eqref{eq:t7} and \eqref{eq:t9} for cases 7) and 9), we have also added the tree level contribution.
For these term the contribution of $\vec p_2$ in $\vec S^+ \cdot (\vec p_1-\vec p_2)$ should also be kept, but in the region of invariant masses of $M_2 B$ that we are interested, it is easy to see that $|\vec p_1|$ is more than an order of magnitude bigger than $|\vec p_2|$ and we disregard $p_2$, which makes the formalism more compact.

Summing coherently the amplitudes for the mechanisms of Figs.~\ref{fig:Fig1a} and \ref{fig:Fig1b}, we arrive at the final formula for the different reactions
\begin{equation}\label{eq:ti}
  t_i = C \bra{\Xi^* (3/2^+)} \vec S^+ \cdot \vec p_i \ket{\Omega_c^0}\; \tilde{t}_i,
\end{equation}
with 
\begin{alignat}{3}\label{eq:tti}
  & \tilde{t}_1=t'_1 +t'_7,        &\quad~~ &\tilde{t}_2=t'_2 +t'_8, \nonumber\\[0.1cm]
  & \tilde{t}_3=t'_3 +t'_9, &\quad~~ &\tilde{t}_4=t'_4 +t'_{10}, \\[0.1cm]  
  & \tilde{t}_5=t'_5, &\quad~~ &\tilde{t}_6=t'_6. \nonumber
\end{alignat}
In Eq.~\eqref{eq:ti}, the baryon $\Xi^* (3/2^+)$ should be the baryon $B$ of the loop, but since the $t_{M_2 B,\, M_i \Xi^*_i}$ is spin independent, the spin of the intermediate $B$ baryon is transfered to the final $\Xi^*$ state, resulting in the formula of Eq.~\eqref{eq:ti}.

Finally, once the $t_i$ matrices have been constructed, the mass distribution for the final $M_i \Xi^*_i$ pair is given by 
\begin{equation}\label{eq:Minv}
  \dfrac{{\rm d} \Gamma_{i}}{{\rm d} M_{\rm inv}(M_i \Xi^*_i)}= \dfrac{1}{(2\pi)^3}\; \dfrac{1}{4 M^2_{\Omega_c}}\; p_{i}\, \tilde{q}_i\; \overline{\sum} \sum|t_i|^2, ~(i=1 \sim  6)
\end{equation}
where $p_i$ is the momentum of the external meson in the weak vertex in the $\Omega_c^0$ rest frame,
and $\tilde{q}_i$ is the momentum of the meson $M_i$ of the final $M_i \Xi^*_i$ pair in the rest frame of that pair.
The magnitude $\overline{\sum} \sum|t_i|^2$, taking into account Eq.~\eqref{eq:SM}, is then given by
\begin{equation}\label{eq:sum}
  \overline{\sum} \sum|t_i|^2= C^2 \,\dfrac{2}{3}\; \vec p_i^{\, 2}\, |\tilde{t}_i|^2.
\end{equation}
We take $C^2=1$ in our calculations.

\section{Results}
\begin{figure}[t]
  \centering
  \includegraphics[scale=0.39]{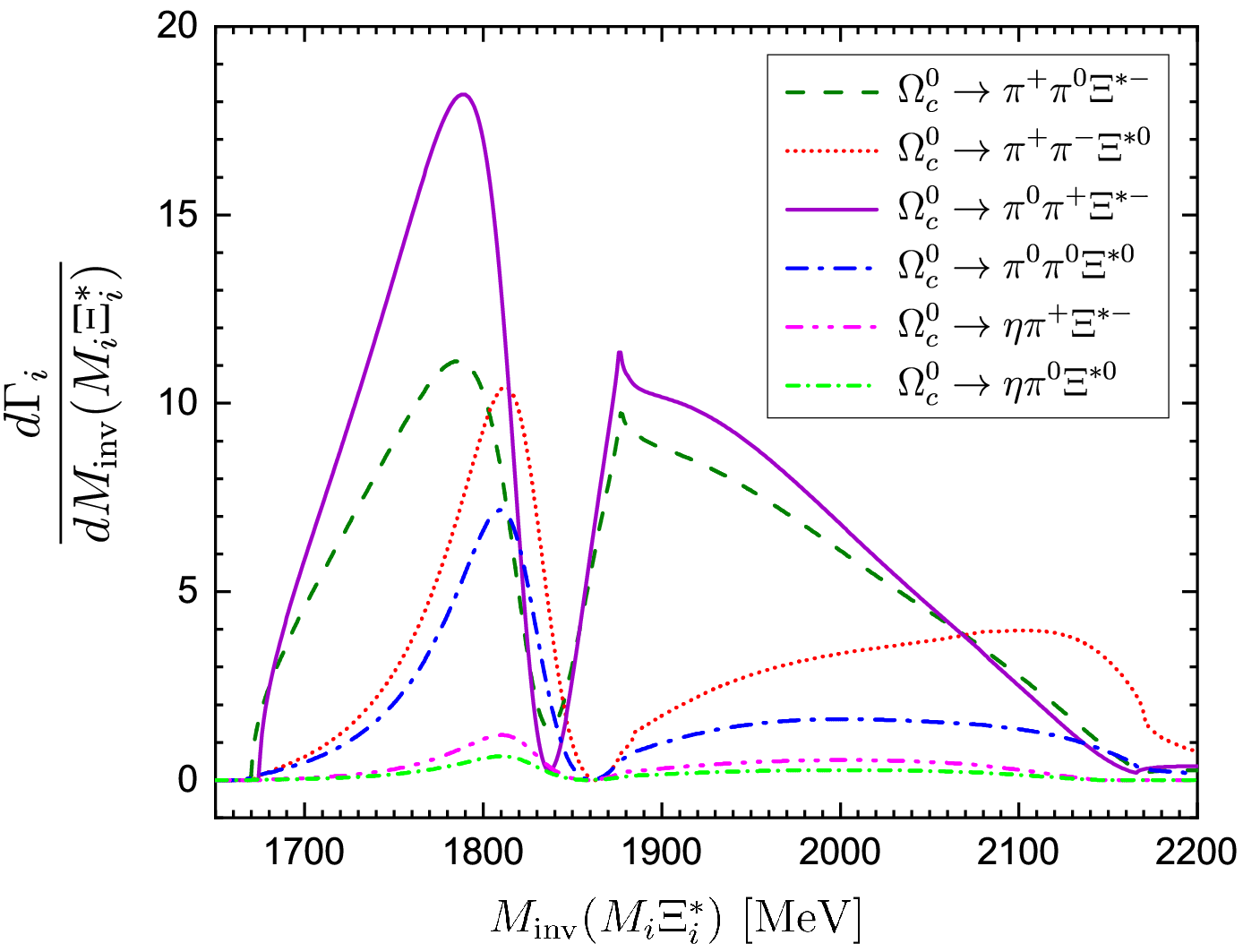}
  \caption{$M_{\rm inv}(M_i \Xi^*_i)$ invariant mass distributions for the $\Omega_c^0 \to M_1 M_i \Xi^*_i$ decays.}
  \label{fig:Fig3}
  \end{figure}

 In Fig.~\ref{fig:Fig3} we show the mass distribution of the final pair for the six reactions that we have studied.

  As we can see, the shapes of the $M_{\rm inv}(M_i \Xi^*_i)$ distributions for the reactions are different from each other, but they share some thing in common: there is a dip in the mass distribution around $1850 \mev$, which is even a zero in all but two of the distributions.
  This is due to the destructive interference of the two resonances, a reminiscence of what happens with the $f_0(500)$ and $f_0(980)$ resonances in $s$-wave $\pi\pi$ scattering, where the $f_0(980)$ shows up in the cross section as a dip, not as a peak.
  This feature does not preclude that the $f_0(980)$ can show up as a peak in many other reactions \cite{review},
  which means that the two $\Xi(1820)$ states can show up also in a different way, as is the case of the $\psi(3686) \to K^- \Lambda \bar \Xi^+$ BESIII reaction \cite{BESIII:2023mlv}.
  The two reactions where the mass distribution does not go to zero are those where the tree level is present.

\begin{figure}[t]
    \centering
    \includegraphics[scale=0.37]{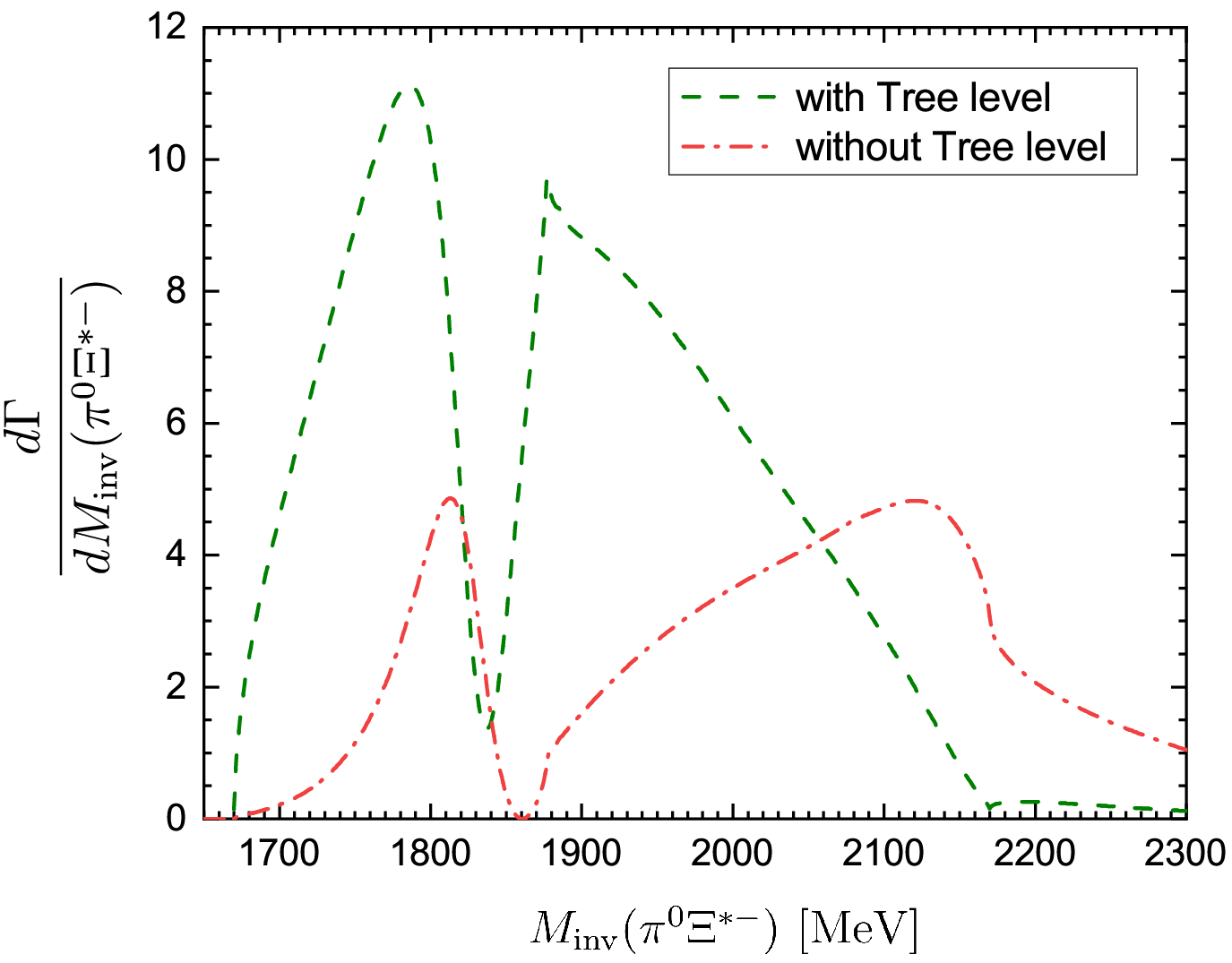}
    \caption{$M_{\rm inv}(\pi^0 \Xi^{*-})$ invariant mass distribution for the $\Omega_c^0 \to \pi^+ \pi^0 \Xi^{*-}$ decay.}
    \label{fig:Fig4}
  \end{figure}

  The mass distributions show two peaks, and it is important not to misidentify them.
  They do not correspond to the two resonances that we are discussing.
  They come from the interference of the two resonance contributions. 
  It is important to notice that the two reactions that contain the tree level contribution, $\Omega_c^0 \to \pi^0 \pi^+ \Xi^{*-}$ and $\Omega_c^0 \to \pi^+ \pi^0 \Xi^{*-}$, have the two peaks more pronounced, and the width of the peaks also do not reflect the widths of the states that we have.
  In order to clarify what happens we show in Fig.~\ref{fig:Fig4} the mass distribution for the $\Omega_c^0 \to \pi^+ \pi^0 \Xi^{*-}$ reaction removing the tree level contribution.
 The peak to the left certainly reflects our first state at $1824 \mev$, with an apparent width of around $40 \mev$, even smaller than the one we get from the pole position, $62\mev$, due to the interference with the second resonance.
 The shape of the second resonance, with a width of $260\mev$ from the pole position, cannot be distinguished due again to the destructive interference with the fist peak, but one can guess that it is a broad resonance, otherwise one could still see a sharper structure than the one we obtain around $1900-2000 \mev$.
 It is interesting to see that around $2150\mev$ there is another peak.
 This corresponds to a third resonance obtained in Ref.~\cite{Molina:2023uko} and also Ref.~\cite{Sarkar:2004jh} around that energy.

 After this discussion it becomes clear that the reactions free of tree level contribution show clearer the resonance structure.
 Coming back to Fig.~\ref{fig:Fig3}, these are the reactions: $\Omega_c^0 \to \pi^+ \pi^- \Xi^{*0}$, $\Omega_c^0 \to \pi^0 \pi^0 \Xi^{*0}$, $\Omega_c^0 \to \eta \pi^+ \Xi^{*-}$, and $\Omega_c^0 \to \eta \pi^0 \Xi^{*0}$.
 The first peak corresponding to the lower $\Xi(1820)$ resonance is clearly seen,
 and the interference pattern is very similar in all the reactions.
 The peak corresponding to a third resonance around $2150\mev$ is better seen in the $\Omega_c^0 \to \pi^+ \pi^- \Xi^{*0}$ reaction.

 One word of caution should be said here.
 If we have the $\Omega_c^0 \to \pi^0 \pi^0 \Xi^{*0}$ reaction, we should have symmetrized our amplitude with respect to the two $\pi^0$ identical states.
 We have not done it because the kinematics of these two $\pi^0$ are very different and one can clearly distinguish them.
 One can see that the external $\pi^0$ coming from the weak vertex has a momentum around $770\mev$ while the $\pi^0$ from the $\pi^0 \Xi^{*0}$ resonance state has about $50\mev$.
 They are easily distinguished experimentally as shown in Ref.~\cite{LHCb:2019tdw}.
 This argument also holds to distinguish the two pions in other reactions into the one coming from the weak vertex and the one belonging to the resonance.

 \section{Conclusions}
 In this work we have studied several reactions coming from the single Cabibbo suppressed weak decay of the $\Omega_c^0$ state into two pseudoscalars and a $\Xi^*$ state. 
 One of the pseudoscalars interacts with the $\Xi^*$ state to produce two resonances around $1820 \mev$, that were predicted in Ref.~\cite{Sarkar:2004jh} and reconfirmed in Ref.~\cite{Molina:2023uko}. 
 These resonances played an important role describing the peak seen in the $K^- \Lambda$ mass distribution in the $\psi(3686)$ decay to $K^- \Lambda \bar \Xi^{*+}$ of the BESIII experiment \cite{BESIII:2023mlv}.

 In this work we study six reactions: 
 $\Omega_c^0 \to \pi^+ \pi^0 \Xi^{*-}$, $\Omega_c^0 \to \pi^+ \pi^- \Xi^{*0}$, $\Omega_c^0 \to \pi^0 \pi^+ \Xi^{*-}$, $\Omega_c^0 \to \pi^0 \pi^0 \Xi^{*0}$, $\Omega_c^0 \to \eta \pi^+ \Xi^{*-}$, and $\Omega_c^0 \to \eta \pi^0 \Xi^{*0}$, 
where the first meson is produced at the weak vertex and the second meson comes from the decay of the resonances. We obtained mass distribution for the final pair which differed from each other in the different reactions. 
In particular the shapes of the $\Omega^0_c \to \pi^+ \pi^0 \Xi^{*-}$ and $\Omega^0_c \to \pi^0 \pi^+ \Xi^{*-}$ were very different to the other ones and this was traced back to the contribution of the tree level mechanism to the reaction. 
The other four reactions did not have tree level contribution and required rescattering of meson baryon where the resonances are produced. 
The shapes of these four reactions resembled each other and had as a distinct feature, very different to the one observed in the $\psi(3686)$ decay to $K^- \Lambda \bar \Xi^{*+}$, which is a dip of the mass distribution around $1850 \mev$, that was due to a destructive interference between the two $\Xi(1820)$ states. 
This pattern reminds one of the same thing happening in the $s$-wave $\pi \pi$ cross section around $980 \mev$, where a dip is also observed as a consequence of the destructive interference between the $f_0(500)$ and the $f_0(980)$ resonances. 

  At present many Cabibbo favored $\Omega^0_c$ decays into strangeness $S=-3$ states have been reported  in the PDG \cite{PDG:2022}, but updates of Belle and LHCb will open the door to the observation of single Cabibbo decay modes, as the one reported here. We are looking forward to these updates, encouraging the performance of the suggested experiments which will shed light into the existence of two close by $\Xi(1820)$ states, and related to it, on the nature of such states. 

  \section*{Acknowledgments}
  R. M. acknowledges support from the CIDEGENT program with Ref. CIDEGENT/2019/015, the Spanish Ministerio de Economia
y Competitividad and European Union (NextGenerationEU/PRTR) by the grant with Ref. CNS2022-136146.
  This work is partly supported by the National Natural Science Foundation of China (NSFC) under Grants No. 12365019 and No. 11975083, and by the Central Government Guidance Funds for Local Scientific and Technological Development, China (No. Guike ZY22096024).
  This work is also partly supported by the Spanish Ministerio de Economia y Competitividad (MINECO) and European FEDER
  funds under Contracts No. FIS2017-84038-C2-1-P B, PID2020-112777GB-I00, and by Generalitat Valenciana under contract
  PROMETEO/2020/023.
  This project has received funding from the European Union Horizon 2020 research and innovation
  programme under the program H2020-INFRAIA-2018-1, grant agreement No. 824093 of the STRONG-2020 project.

\bibliography{refs}

\end{document}